\documentclass[12pt]{article}
\usepackage{amssym}

\renewcommand{\theequation}{\arabic{section}.\arabic{equation}}

\newcommand{\aalpha}{\overline{\alpha}}
\newcommand{\bbeta}{\overline{\beta}}
\def\case#1#2{{\textstyle{#1\over #2}}}
\def\C{\mbox{$\Bbb C$}}

\title{
Harmonic oscillator with nonzero minimal uncertainties in both position and momentum
in a SUSYQM framework}
\author{C Quesne$^{\dagger}$ and  V M Tkachuk$^{\ddagger}$\\
$^{\dagger}$ {\small Physique Nucl\'eaire Th\'eorique et Physique
Math\'ematique,  Universit\'e Libre de Bruxelles,} \\ 
{\small Campus de la Plaine CP229, Boulevard~du Triomphe, B-1050 Brussels,
Belgium}\\ 
$^{\ddagger}$ {\small Ivan Franko Lviv National University, Chair of Theoretical
Physics,}\\
{\small 12, Drahomanov Street, Lviv UA-79005, Ukraine}\\
{\small E-mail: cquesne@ulb.ac.be and tkachuk@ktf.franko.lviv.ua}}
\date{ }
\begin{document}
\baselineskip=20pt plus 1pt minus 1pt
\maketitle
\begin{abstract}
In the context of a two-parameter $(\alpha, \beta)$ deformation of the canonical
commutation relation leading to nonzero minimal uncertainties in both position and
momentum, the harmonic oscillator spectrum and eigenvectors are determined by using
an extension of the techniques of conventional supersymmetric quantum mechanics
combined with shape invariance under parameter scaling. The resulting supersymmetric
partner Hamiltonians  correspond to different masses and frequencies. The exponential
spectrum is proved to reduce to a previously found quadratic spectrum whenever one of
the parameters
$\alpha$, $\beta$ vanishes, in which case  shape invariance under parameter translation
occurs. In the special case where $\alpha = \beta \ne 0$, the oscillator Hamiltonian is
shown to coincide with that of the $q$-deformed oscillator with $q > 1$ and its
eigenvectors are therefore $n$-$q$-boson states. In the general case where $0 \ne
\alpha \ne \beta \ne 0$, the eigenvectors are constructed as linear combinations of
$n$-$q$-boson states by resorting to a Bargmann representation of the latter and to
$q$-differential calculus. They are finally expressed in terms of a $q$-exponential  and
little $q$-Jacobi polynomials. 
\end{abstract}

\vspace{0.5cm}


\newpage
\section{Introduction}

Studies on small distances in string theory and quantum gravity suggest the existence of
a finite lower bound to the possible resolution of length $\Delta x_0$ (see, e.g.,
\cite{gross, maggiore}) On the other hand, on large scales there is no notion of plane
waves or momentum eigenvectors on generic curved spaces. It has therefore been
suggested that there could also exist a finite lower bound to the possible resolution of
momentum $\Delta p_0$ (see, e.g., \cite{kempf94a}). It is a natural, though nontrivial,
assumption that minimal length and momentum should quantum theoretically be described
as nonzero minimal uncertainties in position and momentum measurements.\par
%
%
Such nonzero minimal uncertainties can be described in the framework of small corrections
to the canonical commutation relation~\cite{kempf94b, hinrichsen}
\begin{equation}
  [x, p] = {\rm i} \hbar (1 + \aalpha x^2 + \bbeta p^2)  \label{eq:def-com} 
\end{equation}
with $\aalpha \ge 0$, $\bbeta \ge 0$, and $\aalpha \bbeta < \hbar^{-2}$. In such a
context, they are given by $\Delta x_0 = \hbar \sqrt{\bbeta/(1 - \hbar^2 \aalpha
\bbeta)}$ and $\Delta p_0 = \hbar \sqrt{\aalpha/(1 - \hbar^2 \aalpha \bbeta)}$,
respectively\footnote{In \cite{kempf94b, hinrichsen}, the two deforming parameters are
denoted by $\alpha$ and $\beta$. Here we reserve this notation for the dimensionless
parameters to be introduced in section~2.}.\par
%
%
Since the canonical commutation relations lie in the very heart of quantum mechanics,
studying the influence of small corrections to them in a quantum mechanical framework is
interesting in its own right. It has indeed been argued~\cite{kempf97} that such
corrections may provide an effective description not only of strings but also of
non-pointlike particles such as quasiparticles and various collective excitations in solids,
or composite particles such as nucleons and nuclei.\par
%
%
Solving quantum mechanical problems with the deformed canonical commutation relation
(\ref{eq:def-com}) may, however, be a difficult task. In the special case where $\aalpha =
0$ and $\bbeta > 0$, the minimal uncertainty in the position turns out to be $\Delta x_0
= \hbar \sqrt{\bbeta}$, whereas there is no nonzero minimal momentum uncertainty. As a
consequence, equation (\ref{eq:def-com}) can be represented on momentum space wave
functions (although not on position ones). Similarly, in the case where $\aalpha > 0$ and
$\bbeta = 0$, there is only a nonzero minimal uncertainty in the momentum $\Delta
p_0 = \hbar \sqrt{\aalpha}$ and equation (\ref{eq:def-com}) can be represented on
position space wave functions. On the contrary, in the general case where $\aalpha > 0$
and $\bbeta > 0$, there is neither position nor momentum representation, so that one
has to resort to a generalized Fock space representation or, equivalently,
to the corresponding Bargmann representation~\cite{kempf94b, hinrichsen,
kempf93}.\par
%
%
The investigation of the harmonic oscillator with Hamiltonian
\begin{equation}
  H = \frac{p^2}{2m} + \frac{1}{2} m \omega^2 x^2
\end{equation}
where $x$ and $p$ satisfy the deformed canonical commutation relation
(\ref{eq:def-com}), is an interesting and nontrivial topic. The eigenvalue problem for such
an oscillator
\begin{equation}
  H |\psi_n\rangle = E_n |\psi_n\rangle \qquad n = 0, 1, 2, \ldots  \label{eq:eigen-prob} 
\end{equation}
has been solved exactly only in the case where $\aalpha = 0$ and $\bbeta > 0$ by using
the momentum representation and the technique of differential equations~\cite{kempf95}.
This approach has been recently extended to $D$ dimensions~\cite{chang} and some
ladder operators have been constructed~\cite{dadic}. While it is obvious that in the case
where $\aalpha > 0$ and $\bbeta = 0$, the eigenvalue problem can be treated in a similar
way, that corresponding to both $\aalpha > 0$ and $\bbeta > 0$ is more complicated
and, as far as we know, has not been solved so far.\par
%
%
Over the years, it has been shown that supersymmetric quantum mechanics (SUSYQM)
plays an important role in obtaining exact solutions of quantum mechanical problems
(see, e.g.,~\cite{cooper, junker}). In fact, all solvable quantum mechanical problems are
either supersymmetric or can be made so.\par
%
%
Among the various exactly solvable potentials, there is a certain class of potentials
characterized by a property known as shape invariance~\cite{gendenshtein}. Shape
invariant potentials are potentials such that their SUSY partner has the same spatial
dependence with possibly altered parameters. For such potentials, both the energy
eigenvalues and the wave functions can be obtained algebraically without solving the
differential equation~\cite{gendenshtein, dabrowska}. It turns out that the formalism of
SUSYQM plus shape invariance (connected with translations of parameters) is intimately
related to the factorization method developed by Schr\" odinger~\cite{schrodinger} and
by Infeld and Hull~\cite{infeld} (for a comparison between these two methods and
corresponding references see~\cite{junker}). Other types of shape invariance, such as
that connected with parameter scaling~\cite{spiridonov, khare}, have provided a lot of
new exactly solvable problems.\par
%
%
The purpose of the present paper is to extend SUSYQM and the notion of shape invariance
to eigenvalue problems in the context of the deformed canonical commutation relation
(\ref{eq:def-com}) and to apply such an extension to the case of the harmonic oscillator 
with nonzero minimal uncertainties in both position and momentum (i.e.,
$\aalpha > 0$ and $\bbeta > 0$). As a result, we will derive exact results for the energy
spectrum and the eigenstates of such a system in a purely algebraic way.\par
%
%
This paper is organized as follows. The harmonic oscillator spectrum is obtained in
section~2. Some special cases are reviewed in section~3. In section~4, the Hamiltonian
eigenvectors are determined in explicit form. Section~5 contains the conclusion. Finally,
some mathematical details are to be found in the appendix.\par
%
%
\section{Harmonic oscillator spectrum in the general case}
\setcounter{equation}{0}

It is convenient to introduce dimensionless position and momentum operators, $X = x/a$
and $P = pa/\hbar$, where $a = \sqrt{\hbar/(m\omega)}$ is the oscillator characteristic
length. They satisfy the commutation relation
\begin{equation}
  [X, P] = {\rm i}(1+ \alpha X^2 + \beta P^2)  \label{eq:def-com-bis}
\end{equation}
where $\alpha$ and $\beta$ denote the dimensionless parameters $\alpha = \aalpha
\hbar/(m\omega)$ and $\beta = \bbeta m \hbar \omega$, respectively. Here we have
$\alpha \ge 0$, $\beta \ge 0$, and $\alpha \beta < 1$.\par
%
%
The dimensionless harmonic oscillator Hamiltonian is then given by
\begin{equation}
  h = \frac{H}{\hbar\omega} = \frac{1}{2} (P^2 + X^2)  \label{eq:h}
\end{equation}
and the corresponding eigenvalue problem reads
\begin{equation}
  h |\psi_n\rangle = e_n |\psi_n\rangle \qquad n = 0, 1, 2, \ldots 
\end{equation}
where $e_n = E_n/(\hbar\omega)$. To find the spectrum $e_n$, $n=0$, 1, 2,~\ldots,
we shall proceed in two steps: we will show that the Hamiltonian $h$ is factorizable, then
we will prove that the factorized Hamiltonian satisfies a condition similar to the shape
invariance condition of conventional SUSYQM.\par
%
%
Let us first try to write $h$ in the factorized form
\begin{equation}
  h = B^+(g,s) B^-(g,s) + \epsilon_0  \label{eq:fact-h}
\end{equation}
where \begin{equation}
  B^{\pm}(g,s) = \frac{1}{\sqrt{2}} (s X \mp {\rm i} g P)  \label{eq:B+/-}
\end{equation}
and $\epsilon_0$ is the factorization energy. In (\ref{eq:B+/-}), $g$ and $s$ are
assumed to be two positive constants that are some functions of $\alpha$, $\beta$
and which go to 1 in the limit $\alpha$, $\beta \to 0$. The operators $B^+(g,s)$ and
$B^-(g,s)$ are therefore counterparts of the standard harmonic oscillator creation and
annihilation operators, $a^+ = (X - {\rm i}P)/\sqrt{2}$ and $a = (X + {\rm i}P)/\sqrt{2}$,
respectively.\par
%
%
Inserting (\ref{eq:B+/-}) in (\ref{eq:fact-h}) leads to the equation
\begin{equation}
  h = \case{1}{2} \left[(g^2 - \beta g s) P^2 + (s^2 - \alpha g s) X^2 - g s\right] + 
  \epsilon_0. 
\end{equation}
For this expression of $h$ to be equivalent to that given in (\ref{eq:h}), three conditions
have to be fulfilled, namely
\begin{eqnarray}
  g^2 - \beta g s & = & 1  \label{eq:fact-cond1} \\
  s^2 - \alpha g s & = & 1  \label{eq:fact-cond2} \\
  \epsilon_0 & = & \case{1}{2} g s.  \label{eq:fact-cond3}
\end{eqnarray} 
\par
%
%
It can be easily shown that equations (\ref{eq:fact-cond1}) and (\ref{eq:fact-cond2})
admit positive solutions for $g$ and $s$, given by
\begin{equation}
  g = sk \qquad s = \frac{1}{\sqrt{1 - \alpha k}}  \label{eq:g-s}
\end{equation}
where
\begin{equation}
  k \equiv \case{1}{2}(\beta - \alpha) + \sqrt{1 + \case{1}{4}(\beta - \alpha)^2}.
  \label{eq:k}
\end{equation}
In deriving the expression of $s$ in (\ref{eq:g-s}), use has been made of the equivalence
between the conditions $1 - \alpha k > 0$ and $\alpha \beta < 1$. Furthermore, one
checks that $k$, $g$, $s \to 1$ for $\alpha$, $\beta \to 0$, as it should be. We
conclude that $h$ can be written in the form (\ref{eq:fact-h}), where the factorization
energy
$\epsilon_0$ is given in (\ref{eq:fact-cond3}) and the two parameters $\alpha$, $\beta$
of the problem have been replaced by their combinations $g$, $s$, defined in
(\ref{eq:g-s}) and (\ref{eq:k}).\par
%
%
Let us now consider a hierarchy of Hamiltonians
\begin{equation}
  h_i = B^+(g_i, s_i) B^-(g_i, s_i) + \sum_{j=0}^i \epsilon_j \qquad  i=0, 1, 2, \ldots
  \label{eq:hierarchy}
\end{equation}
whose first member coincides with (\ref{eq:fact-h}), i.e., $h_0 = h$. Here $g_i$, $s_i$,
$\epsilon_i$, $i=1$, 2,~\ldots are assumed to be some positive parameters and $g_0 =
g$, $s_0 = s$. We shall now proceed to prove that one can find values of $g_i$,
$s_i$, $\epsilon_i$, $i=1$, 2,~\ldots, such that the condition
\begin{equation}
  B^-(g_i, s_i) B^+(g_i, s_i) = B^+(g_{i+1}, s_{i+1}) B^-(g_{i+1}, s_{i+1}) 
  + \epsilon_{i+1}  \label{eq:SI}
\end{equation}
is satisfied. Equation (\ref{eq:SI}), written in operator form, is exactly equivalent to the
equation used in the factorization method~\cite{schrodinger, infeld} (see
also~\cite{green}) and is similar to the shape invariance condition of conventional
SUSYQM~\cite{cooper, junker, gendenshtein}. Note that the factorization method (and
thus the SUSY one) is quite general (see, e.g.,~\cite{green}) and can be used for finding
the eigenvalues of arbitrary Hermitian operators with a bounded-from-below spectrum. As
it is obvious that the eigenvalues of the Hamiltonian (\ref{eq:h}) are positive, we can
apply the factorization method or the SUSY one with shape invariance to our problem.\par
%
%
In explicit form, equation (\ref{eq:SI}) reads
\begin{eqnarray}
  \lefteqn{\case{1}{2} \left[(g_i^2 + \beta g_i s_i) P^2 + (s_i^2 + \alpha g_i s_i) X^2 +
        g_i s_i\right] \nonumber } \\
  & = & \case{1}{2} \left[(g_{i+1}^2 - \beta g_{i+1} s_{i+1}) P^2 + (s_{i+1}^2 - \alpha
        g_{i+1} s_{i+1}) X^2 - g_{i+1} s_{i+1}\right] + \epsilon_{i+1} 
\end{eqnarray}
where $i=0$, 1, 2,~\ldots. This leads to the set of three relations
\begin{eqnarray}
  g_{i+1}^2 - \beta g_{i+1} s_{i+1} & = & g_i^2 + \beta g_i s_i  \label{eq:SI-cond1} \\
  s_{i+1}^2 - \alpha g_{i+1} s_{i+1} & = & s_i^2 + \alpha g_i s_i  \label{eq:SI-cond2} \\
  \epsilon_{i+1} & = & \case{1}{2}(g_i s_i + g_{i+1} s_{i+1}).  \label{eq:SI-cond3}
\end{eqnarray}
\par
%
%
At this stage, it is worth noting that by multiplying (\ref{eq:SI-cond1}) by $\alpha$ and
(\ref{eq:SI-cond2}) by $\beta$, then subtracting, we get the equation
\begin{equation}
  g_{i+1}^2 - \gamma^2 s_{i+1}^2 = g_i^2 - \gamma^2 s_i^2  \label{eq:hyperbola}
\end{equation}
where 
\begin{equation}
  \gamma \equiv \sqrt{\frac{\beta}{\alpha}}.  \label{eq:gamma}
\end{equation}
On iterating (\ref{eq:hyperbola}) and using (\ref{eq:fact-cond1}) and
(\ref{eq:fact-cond2}), it is then obvious that the recursion for $g_i$, $s_i$ runs along
the arc of the hyperbola $g^2 - \gamma^2 s^2 = 1 - \gamma^2$ that lies in the first
quadrant of the $(g,s)$ plane. Since the equation of this hyperbola can also be written as
$uv = 1 - \gamma^2$, where $u \equiv g + \gamma s = 0$ and $v \equiv g - \gamma s
= 0$ are the equations of the asymptotes, it may prove useful to replace $(g,s)$ by
$(u,v)$.\par
%
%
Let us therefore introduce the new combinations of parameters
\begin{equation}
  u_i = g_i + \gamma s_i \qquad v_i = g_i - \gamma s_i.
\end{equation}
The inverse transformation reads
\begin{equation}
  g_i = \frac{1}{2}(u_i + v_i) \qquad s_i = \frac{1}{2\gamma}(u_i - v_i) 
  \label{eq:gs-uv}
\end{equation}
where the assumptions $g_i$, $s_i > 0$ impose that $u_i > |v_i|$. On substituting
(\ref{eq:gs-uv}) into (\ref{eq:SI-cond1}), (\ref{eq:SI-cond2}), then combining the two
resulting equations, we obtain the relations
\begin{eqnarray}
  u_{i+1}^2 + q v_{i+1}^2 & = & v_i^2 + q u_i^2  \label{eq:SI-cond1a} \\
  u_{i+1} v_{i+1} & = & u_i v_i  \label{eq:SI-cond2a}
\end{eqnarray}
where 
\begin{equation}
  q \equiv \frac{1 + \sqrt{\alpha\beta}}{1 - \sqrt{\alpha\beta}} > 1  \label{eq:q}
\end{equation}
and equation (\ref{eq:SI-cond2a}) coincides with (\ref{eq:hyperbola}).
\par
%
%
Equations (\ref{eq:SI-cond1a}) and (\ref{eq:SI-cond2a}) suggest a further transformation
from $u_i$, $v_i$ to
\begin{equation}
  d_i = u_i v_i \qquad t_i = \frac{v_i}{u_i}  \label{eq:d-t}
\end{equation}
where $d_i$ and $t_i$ have the same sign, which is that of $v_i$, and $|t_i| < 1$.
According to (\ref{eq:SI-cond2a}), $d_i$ is actually independent of $i$,
\begin{equation}
  d_i = d = uv  \label{eq:SI-cond1b}
\end{equation}
which amounts to the hyperbola equation previously obtained. On the other hand,
equation (\ref{eq:SI-cond1a}) can be rewritten as
\begin{equation}
  q t_{i+1} - t_i = \frac{q t_{i+1} - t_i}{t_i t_{i+1}}
\end{equation}
thus showing that $q t_{i+1} - t_i = 0$ or 
\begin{equation}
  t_i = q^{-i} t \qquad t \equiv \frac{v}{u} = \frac{k - \gamma}{k + \gamma}.
  \label{eq:SI-cond2b}
\end{equation}
From (\ref{eq:d-t}), (\ref{eq:SI-cond1b}), and (\ref{eq:SI-cond2b}), we also obtain
\begin{equation}
  u_i = q^{i/2} u \qquad v_i = q^{-i/2} v.  \label{eq:SI-cond12c}
\end{equation}
We conclude that the extended shape invariance condition (\ref{eq:SI}) can indeed be
satisfied by keeping the combination of parameters $d$ constant while scaling the other
combination of parameters $t$ according to equation (\ref{eq:SI-cond2b}). Note that
for $i \to \infty$, $v_i \to 0$, so that the recursion actually reaches the hyperbola
asymptote lying in the first quadrant of the $(g,s)$ plane.\par
%
%
The eigenvalues $e_n$ of $h$ are therefore given by
\begin{eqnarray}
  e_n(q,t) & = & \sum_{i=0}^n \epsilon_i = \sum_{i=0}^{n-1} g_i s_i + \frac{1}{2} g_n
         s_n = \frac{1}{4\gamma} \left(\sum_{i=0}^{n-1} (u_i^2 - v_i^2) + \frac{1}{2}
         (u_n^2 - v_n^2)\right) \nonumber \\
  & = & \frac{1}{4\gamma} \left\{u^2 \left([n]_q + \frac{1}{2} q^n\right) - v^2 
         \left([n]_{q^{-1}} + \frac{1}{2} q^{-n}\right)\right\} \nonumber \\
  & = & \frac{1}{4\gamma} \left\{\left(u^2 - \frac{v^2}{q^{n-1}}\right) [n]_q + 
         \frac{1}{2} \left(u^2 q^n - \frac{v^2}{q^n}\right)\right\} \nonumber \\
  & = & \frac{u^2}{4\gamma} \left\{\left(1 - \frac{t^2}{q^{n-1}}\right) [n]_q + 
         \frac{1}{2} \left(q^n - \frac{t^2}{q^n}\right)\right\}  \label{eq:energy}
\end{eqnarray}
where we have successively used equations (\ref{eq:fact-cond3}), (\ref{eq:SI-cond3}),
(\ref{eq:gs-uv}), (\ref{eq:SI-cond12c}), (\ref{eq:SI-cond2b}), and the definitions
\begin{equation}
  [n]_q \equiv \frac{q^n-1}{q-1} \qquad [n]_{q^{-1}} \equiv \frac{q^{-n}-1}{q^{-1}-1}
  = q^{-n+1} [n]_q.  \label{eq:q-number} 
\end{equation}
In (\ref{eq:energy}), we employ the notation $e_n(q,t)$ to stress that, apart from a
multiplicative constant $u^2/(4\gamma)$, the energy eigenvalues depend on the two
parameters $q$ and $t$, defined in (\ref{eq:q}) and (\ref{eq:SI-cond2b}),
respectively.\par
%
%
It can be easily shown that $q(\alpha, \beta) = q(\beta, \alpha)$, $u^2(\alpha,
\beta)/[4\gamma(\alpha, \beta)] = u^2(\beta, \alpha)/[4\gamma(\beta, \alpha)]$,
and $t(\alpha, \beta) = - t(\beta, \alpha)$, where we explicitly write down the
dependence on the deformation parameters $\alpha$, $\beta$. As a result, the
eigenvalues $e_n$ are symmetric under exchange of $\alpha$ and $\beta$.\par
%
%
{}From (\ref{eq:energy}), we obtain for the ground state and excitation energies
\begin{equation}
  e_0(q,t) = \frac{u^2}{8\gamma} (1-t^2)
\end{equation}
and
\begin{equation}
  e_n(q,t) - e_0(q,t) = \frac{1}{2} K^2 \left(1 - \frac{t^2}{q^n}\right) [n]_q \qquad
  n= 1, 2, \ldots  \label{eq:exc-energy}
\end{equation}
where
\begin{equation}
  K \equiv u \sqrt{\frac{q+1}{4\gamma}}.  \label{eq:K}
\end{equation}
Since $q > 1$ and $t^2 < 1$, the excitation energies grow exponentially to infinity
when $n \to \infty$. Such a feature has already been encountered before in SUSYQM and
shape invariance associated with parameter scaling~\cite{spiridonov}.\par
%
%
On using (\ref{eq:B+/-}), (\ref{eq:gs-uv}), (\ref{eq:SI-cond2b}), and
(\ref{eq:SI-cond12c}), the Hamiltonians (\ref{eq:hierarchy}) of the SUSYQM hierarchy can
be written as
\begin{equation}
  h_i = \case{1}{2}(a_i P^2 + b_i X^2) + c_i \qquad i=0, 1, 2, \ldots
\end{equation}
where
\begin{eqnarray}
  a_i & = & \frac{u^2}{2(q+1)} (q^i + t) \left(1 + \frac{t}{q^{i-1}}\right) \nonumber \\
  b_i & = & \frac{u^2}{2 \gamma^2 (q+1)} (q^i - t) \left(1 - \frac{t}{q^{i-1}}\right) \\
  c_i & = & \frac{u^2}{4\gamma} \left(1 - \frac{t}{q^{i-1}}\right) [i]_q. \nonumber 
\end{eqnarray}
For the supersymmetric partner of $h = h_0$, for instance, we obtain
\begin{equation}
  h_1 = \frac{1}{1 - \alpha k} \left\{\frac{1}{2} [1 + (2\beta - \alpha)k] P^2 +
  \frac{1}{2} (1 + \alpha k) X^2 + k\right\}
\end{equation}
where $k$ is defined in (\ref{eq:k}). Going back to variables with dimensions, we get
\begin{equation}
  H_i \equiv \hbar \omega h_i = \frac{p^2}{2m_i} + \frac{1}{2} m_i \omega_i^2 x^2
  + c_i \hbar \omega  \label{eq:hierarchy-bis}
\end{equation}
with
\begin{equation}
  m_i = \frac{m}{a_i} \qquad \omega_i = \sqrt{a_i b_i}\, \omega.  \label{eq:mass}
\end{equation}
\par
%
%
We conclude that the harmonic oscillator with nonzero minimal uncertainties in both
position and momentum belongs to a hierarchy of Hamiltonians of the same type but with
different masses and frequencies. The change of the kinetic energy term, appearing
naturally in addition to the usual modification of the potential energy one, is a new
feature arising from the deformation of the canonical commutation relation. It is distinct
from the supersymmetric generation of combined potential and effective-mass variations
that may be effected for Hamiltonians with both a position-dependent effective mass
and a position-dependent potential in the context of conventional
SUSYQM~\cite{milanovic}.\par
%
%
\section{Some special cases}
\setcounter{equation}{0}

In the present section, we will examine the two special cases where one of the
parameters, e.g., $\alpha$, vanishes or both parameters $\alpha$, $\beta$ are equal.\par
%
%
\subsection{\boldmath Limit $\alpha \to 0$}

We plan to show that in the limit $\alpha \to 0$, the energy spectrum (\ref{eq:energy})
reproduces the results previously obtained in the momentum
representation~\cite{kempf95, chang}.\par
%
%
{}For small $\alpha$ values, the parameter $q$ behaves as
\begin{equation}
  q \simeq 1 + 2 \sqrt{\alpha\beta} + O(\alpha)
\end{equation}
so that
\begin{equation}
  q^n \simeq 1 + 2n \sqrt{\alpha\beta} + O(\alpha) \qquad [n]_q \simeq n + 
  O(\sqrt{\alpha}).
\end{equation}
Furthermore,
\begin{eqnarray}
  \frac{1}{4\gamma}\left(u^2 - \frac{v^2}{q^{n-1}}\right) & \simeq & gs + \frac{1}{2}
         \beta s^2 (n-1) + O(\sqrt{\alpha}) \nonumber \\
  \frac{1}{8\gamma}\left(u^2 q^n - \frac{v^2}{q^n}\right) & \simeq & \frac{1}{2} gs +
         \frac{1}{2} \beta s^2 n + O(\sqrt{\alpha}).
\end{eqnarray}
Inserting such results in (\ref{eq:energy}), we get
\begin{equation}
  e_n \simeq gs \left(n + \case{1}{2}\right) + \case{1}{2} \beta s^2 n^2 + 
  O(\sqrt{\alpha})
\end{equation}
where
\begin{equation}
  g \simeq \case{1}{2} \beta + \sqrt{1 + \case{1}{4} \beta^2} + O(\alpha) \qquad
  s \simeq 1 + O(\alpha).
\end{equation}
\par
%
%
In the limit $\alpha \to 0$, we therefore obtain
\begin{equation}
  e_n(\beta) = \left(n + \case{1}{2}\right) \sqrt{1 + \case{1}{4} \beta^2} +
  \case{1}{2} \beta \left(n^2 + n + \case{1}{2}\right)  \label{eq:energy1}
\end{equation}
in agreement with~\cite{kempf95, chang}. We conclude that in such a limit, the
exponential spectrum of the general case reduces to a quadratic one.\par
%
%
It is worth noting that the spectrum corresponding to $\alpha=0$ may also be derived
directly from SUSYQM and shape invariance without resorting to a limiting procedure.
Going back to the factorization conditions (\ref{eq:fact-cond1}) -- (\ref{eq:fact-cond3})
and setting $\alpha=0$ therein, we are only left with a single parameter $g =
\frac{1}{2} \beta + \sqrt{1 + \frac{1}{4} \beta^2}$ since $s=1$. Among the shape
invariance conditions (\ref{eq:SI-cond1}) -- (\ref{eq:SI-cond3}), only the first and third
ones survive, namely
\begin{equation}
  g_{i+1} (g_{i+1} - \beta) = g_i (g_i + \beta) \qquad \epsilon_{i+1} = \case{1}{2}
  (g_i + g_{i+1}).
\end{equation}
The solution of the former is $g_{i+1} = g_i + \beta$, thus giving $g_i = g + i \beta$,
while the latter directly leads to the energy spectrum (\ref{eq:energy1}) by using the
relation $e_n(\beta) = \sum_{i=0}^n \epsilon_i$.\par
%
%
This shows that the harmonic oscillator with only a nonzero minimal uncertainty in the
position is shape invariant under translation of the parameter $g$. The Hamiltonians
$h_i$ of the SUSYQM hierarchy now reduce to
\begin{equation}
  h_i = \case{1}{2} \left[\left(1 + i^2 \beta^2 + 2i \beta \sqrt{1+ \case{1}{4} \beta^2}
  \right) P^2 + X^2 + i \left(i \beta + 2 \sqrt{1+ \case{1}{4} \beta^2}\right)\right]
  \qquad i=0, 1, 2, \ldots.
\end{equation}
From (\ref{eq:hierarchy-bis}) and (\ref{eq:mass}), it follows that in this special case $m_i
\omega_i^2 = m \omega^2$, so that only the kinetic energy term in the Hamiltonian
$H_i$ is changed.\par
%
%
\subsection{\boldmath Case $\alpha = \beta \ne 0$}

There also occurs a simplification in the general formula (\ref{eq:energy}) for the energy
spectrum whenever the dimensionless parameters $\alpha$ and $\beta$ are equal, which
means that the original parameters $\aalpha$ and $\bbeta$ are related through $\aalpha
= m^2 \omega^2 \bbeta$. In such a case, we find
\begin{equation}
  \gamma = k = 1 \qquad g = s = \frac{1}{\sqrt{1 - \alpha}} \qquad q = \frac{1 +
  \alpha}{1 - \alpha}  \label{eq:gamma-k}
\end{equation}
and thus
\begin{equation}
  u = 2g \qquad v = t = 0.
\end{equation}
After some simple transformation, equation (\ref{eq:energy}) can be written as
\begin{equation}
  e_n(q) = \case{1}{4} (q+1) \{(q+1) [n]_q + 1\} = \case{1}{4} (q+1) ([n]_q +
  [n+1]_q).  \label{eq:energy2}
\end{equation}
\par
%
%
On the other hand, the Hamiltonians $h_i$ of the SUSYQM hierarchy now become
\begin{equation}
  h_i = \case{1}{2} \left\{q^i(P^2 + X^2) + (q+1) [i]_q\right\}.
\end{equation}
From (\ref{eq:hierarchy-bis}) and (\ref{eq:mass}), it follows that in $H_i$ both the mass
and the frequency are scaled according to $m_i = q^{-i} m$, $\omega_i = q^i
\omega$.\par
%
%
The energy spectrum (\ref{eq:energy2}) is similar to that of the $q$-deformed harmonic
oscillator
\begin{equation}
  h_{\rm osc} = \case{1}{4} (q+1) \{b, b^+\}  \label{eq:h-osc}
\end{equation}
where $b^+$ and $b$ are $q$-deformed boson creation and annihilation operators
satisfying the relation~\cite{arik, cq}
\begin{equation}
  b b^+ - q b^+ b = I.  \label{eq:q-com}
\end{equation}
Such operators can indeed be used to construct $n$-$q$-boson states
\begin{equation}
  |n\rangle_q = \frac{(b^+)^n}{\sqrt{[n]_q!}}\, |0\rangle_q \qquad n=0, 1, 2, \ldots
  \label{eq:n-boson}
\end{equation}
spanning an orthonormal basis of a deformed Fock space ${\cal F}_q$, i.e.,
\begin{equation}
  {}_q\langle n'|n \rangle_q = \delta_{n',n} \qquad \sum_{n=0}^{\infty} |n\rangle_q\,
  {}_q \langle n| = I.
\end{equation}
In (\ref{eq:n-boson}), $|0\rangle_q$ is the normalized vacuum state, i.e., 
$b |0\rangle_q = 0$ and ${}_q \langle 0|0 \rangle_q = 1$, and the $q$-factorial
\begin{equation}
  [n]_q!  \equiv  \left\{\begin{array}{ll}
        1 & {\rm if\ } n=0 \\[0.2cm]
        [n]_q [n-1]_q \cdots [1]_q & {\rm if\ } n=1, 2, \ldots
     \end{array}\right.  \label{eq:q-fac}
\end{equation}
is defined in terms of the $q$-numbers $[n]_q$ introduced in (\ref{eq:q-number}). Since
\begin{equation}
  b^+ |n\rangle_q = \sqrt{[n+1]_q}\, |n+1\rangle_q \qquad b |n\rangle_q =
  \sqrt{[n]_q}\, |n-1\rangle_q  
\end{equation}
the $n$-$q$-boson states $|n\rangle_q$ turn out to be the eigenvectors of $h_{\rm
osc}$ with eigenvalues $e_n(q)$, i.e., $h_{\rm osc} |n\rangle_q = e_n(q)
|n\rangle_q$, $n=0$, 1, 2,~\ldots.\par
%
%
This means that in the $\alpha = \beta \ne 0$ case, the harmonic oscillator Hamiltonian
(\ref{eq:h}) with nonzero minimal uncertainties in position and momentum must be
reducible to the $q$-deformed oscillator Hamiltonian (\ref{eq:h-osc}). Such an assertion
is easily proved by setting
\begin{equation}
  X = \case{1}{2} \sqrt{q+1}\, (b^+ + b) \qquad P = \case{\rm i}{2} \sqrt{q+1}\, (b^+ -
  b)
\end{equation}
or, conversely,
\begin{equation}
  b^+ = \frac{1}{\sqrt{q+1}} (X - {\rm i}P) \qquad b = \frac{1}{\sqrt{q+1}} (X + {\rm
  i}P).  \label{eq:b+/b}
\end{equation}
It can indeed be shown from the commutation relation (\ref{eq:def-com-bis}) that $b^+$
and $b$, as defined in (\ref{eq:b+/b}), fulfil the $q$-commutation relation
(\ref{eq:q-com}) and are such that $h = h_{\rm osc}$.\par
%
%
\section{Harmonic oscillator eigenvectors in the general case}
\setcounter{equation}{0}

To construct the eigenvectors of the harmonic oscillator Hamiltonian (\ref{eq:h}) in the
case where neither $\alpha$ nor $\beta$ vanishes, one has to resort to a generalized
Fock space representation~\cite{kempf94b, hinrichsen, kempf93}, wherein
\begin{equation}
  X = \frac{1}{2} \sqrt{\gamma(q+1)}\, (b^+ + b) \qquad P = \frac{\rm i}{2}
  \sqrt{\frac{q+1}{\gamma}}\, (b^+ - b)  \label{eq:XP-b}
\end{equation}
are represented in terms of creation and annihilation operators
\begin{equation}
  b^+ = \frac{1}{\sqrt{q+1}} \left(\frac{1}{\sqrt{\gamma}}\,X - {\rm i}
  \sqrt{\gamma}\,P\right) \qquad b = \frac{1}{\sqrt{q+1}}
  \left(\frac{1}{\sqrt{\gamma}}\,X + {\rm i} \sqrt{\gamma}\,P\right)
  \label{eq:b+/b-gen}
\end{equation}
satisfying the $q$-commutation relation (\ref{eq:q-com}). It is worth noting that in the
special case where $\alpha = \beta$, considered in section 3.2, the operators
(\ref{eq:b+/b-gen}) reduce to the operators (\ref{eq:b+/b}) as a consequence of
equation (\ref{eq:gamma-k}). In the general case where $\alpha \ne \beta$, however, the
Hamiltonians $h$ and $h_{\rm osc}$, defined in (\ref{eq:h}) and (\ref{eq:h-osc}),
respectively, do not coincide any more since we have instead the relation
\begin{equation}
  h = \frac{1}{8} (q+1) \left(\biggl(\gamma - \frac{1}{\gamma}\biggr) [(b^+)^2 + b^2]
  + \biggl(\gamma + \frac{1}{\gamma}\biggr) \{b, b^+\}\right). 
\end{equation}
Hence the eigenvectors of $h$ are some
linear combinations of the $n$-$q$-boson states defined in (\ref{eq:n-boson}).\par
%
%
SUSYQM and shape invariance provide us with some prescriptions to construct such linear
combinations~\cite{cooper, junker, dabrowska}. The ground state is indeed the
normalized state annihilated by the operator $B^-(g,s)$, defined in (\ref{eq:B+/-}), which
we shall rewrite here as $B^-(q,t)$, i.e.,
\begin{equation}
  B^-(q,t) |\psi_0(q,t)\rangle = 0 \qquad \langle \psi_0(q,t)|\psi_0(q,t)\rangle = 1
  \label{eq:gs-cond}
\end{equation}
while the normalized excited states can be determined recursively through the equations
\begin{equation}
  |\psi_{n+1}(q,t)\rangle = [e_{n+1}(q,t) - e_0(q,t)]^{-1/2} B^+(q,t) 
  |\psi_n(q,t_1)\rangle \qquad n=0, 1, 2, \ldots  \label{eq:es-cond}
\end{equation}
where, according to (\ref{eq:SI-cond2b}), $t_1 = t/q$.\par
%
%
On using (\ref{eq:gs-uv}), (\ref{eq:SI-cond2b}), (\ref{eq:SI-cond12c}), and
(\ref{eq:XP-b}), the operators $B^{\pm}(q,t)$ can be written as linear combinations of
$b^+$ and $b$,
\begin{eqnarray}
  B^+(q,t) & = & \sqrt{\frac{q+1}{8\gamma}} (u b^+ - v b) = \frac{1}{\sqrt{2}} K (b^+
        - t b) \\
  B^-(q,t) & = & \sqrt{\frac{q+1}{8\gamma}} (u b - v b^+) = \frac{1}{\sqrt{2}} K (b
        - t b^+)
\end{eqnarray}
where $K$ is given in (\ref{eq:K}).\par
%
%
To work out the explicit form of $|\psi_n(q,t)\rangle$, $n=0$, 1, 2,~\ldots, it is
convenient to use the ($q$-deformed) Bargmann representation of the $q$-boson
operators $b^+$, $b$, associated with the corresponding $q$-deformed coherent
states~\cite{cq}. In such a representation the $n$-$q$-boson states $|n\rangle_q$ are
represented by the functions
\begin{equation}
  \varphi_n(q;\xi) = \frac{\xi^n}{\sqrt{[n]_q!}}  \label{eq:phi-n} 
\end{equation}
so that any vector
$|\psi\rangle_q = \sum_{n=0}^{\infty} c_n(q) |n\rangle_q \in {\cal F}_q$ is realized by
the entire function $\psi(q;\xi) = \sum_{n=0}^{\infty} c_n(q) \varphi_n(q;\xi)$,
belonging to a $q$-deformed Bargmann space ${\cal B}_q$, whose scalar
product has been given in~\cite{cq}. The operators $b^+$ and $b$ become the operator
of multiplication by a complex number $\xi$ and the $q$-differential operator ${\cal
D}_q$, defined by 
\begin{equation}
  {\cal D}_q \psi(q;\xi) = \frac{\psi(q;q\xi) - \psi(q;\xi)}{(q-1)\xi} \label{eq:D-q}
\end{equation}
respectively. Hence, $B^{\pm}(q,t)$ are represented by 
\begin{equation}
  {\cal B}^+(q,t) = \frac{1}{\sqrt{2}} K (\xi - t {\cal D}_q) \qquad {\cal B}^-(q,t) =
  \frac{1}{\sqrt{2}} K ({\cal D}_q - t \xi).  \label{eq:barg-B}
\end{equation}
\par
%
%
The first equation in (\ref{eq:gs-cond}) can therefore be rewritten as
\begin{equation}
  ({\cal D}_q - t \xi) \psi_0(q,t;\xi) = 0.  \label{eq:gs-eq}
\end{equation}
As detailed in the appendix, the solution of this first-order $q$-difference equation can be
easily obtained as
\begin{equation}
  \psi_0(q,t;\xi) = {\cal N}_0(q,t) E_{q^2}\left(\frac{t}{q+1} \xi^2\right)  \label{eq:gs}
\end{equation}
where the $q$-exponential $E_q(\xi)$ is defined by~\cite{exton}
\begin{equation}
  E_q(\xi) = \sum_{n=0}^{\infty} \frac{\xi^n}{[n]_q!}  \label{eq:q-exp}
\end{equation}
and ${\cal N}_0(q,t)$ is some normalization coefficient, determined by the second
condition in equation (\ref{eq:gs-cond}).\par
%
%
{}From (\ref{eq:phi-n}) and (\ref{eq:q-exp}), it follows that the ground state Bargmann
wave function can also be written as
\begin{equation}
  \psi_0(q,t;\xi) = {\cal N}_0(q,t) \sum_{\nu=0}^{\infty} \left(\frac{[2\nu-1]_q!!}
  {[2\nu]_q!!}\right)^{1/2} t^{\nu} \varphi_{2\nu}(q;\xi)  \label{eq:gs-bis}
\end{equation}
where
\begin{equation}
  [2\nu-1]_q!!  \equiv  \left\{\begin{array}{ll}
        1 & {\rm if\ } \nu=0 \\[0.2cm]
        [2\nu-1]_q [2\nu-3]_q \cdots [1]_q & {\rm if\ } \nu=1, 2, \ldots
     \end{array}\right.  \label{eq:q-d-fac1}
\end{equation}
and
\begin{equation}
  [2\nu]_q!!  \equiv  \left\{\begin{array}{ll}
        1 & {\rm if\ } \nu=0 \\[0.2cm]
        [2\nu]_q [2\nu-2]_q \cdots [2]_q & {\rm if\ } \nu=1, 2, \ldots
     \end{array}\right. .  \label{eq:q-d-fac2}
\end{equation}
The ground state is therefore a superposition of all even-$n$-$q$-boson states.\par
%
%
The orthonormality of the functions $\varphi_n(q;\xi)$ in ${\cal B}_q$ implies that the
normalization coefficient in (\ref{eq:gs}) and (\ref{eq:gs-bis}) is given by
\begin{equation}
  {\cal N}_0(q,t) = \left(\sum_{\nu=0}^{\infty} \frac{[2\nu-1]_q!!}{[2\nu]_q!!}
  t^{2\nu}\right)^{-1/2} = [{}_1\phi_0(q; \mbox{---}; q^2, t^2)]^{-1/2}.  
  \label{eq:gs-norm}
\end{equation}
Here ${}_r\phi_s(a_1, a_2, \ldots, a_r; b_1, b_2, \ldots, b_s; q, z )$ denotes a basic
hypergeometric function (see~\cite{gasper} and the appendix). It can be easily
checked that the series in (\ref{eq:gs-norm}) is convergent for all allowed $t^2$ values,
hence showing that the ground state eigenvector is normalizable as it should be.\par
%
%
Considering next the excited states, we find that in Bargmann representation, the
recursion relation (\ref{eq:es-cond}) becomes the equation
\begin{equation}
  \psi_{n+1}(q,t;\xi) = \left\{[n+1]_q \left(1 -
  \frac{t^2}{q^{n+1}}\right)\right\}^{-1/2} (\xi - t {\cal D}_q) \psi_n(q,t_1;\xi) 
  \label{eq:es-eq}
\end{equation}
where use has been made of equations (\ref{eq:exc-energy}) and (\ref{eq:barg-B}). As
shown in the appendix, the solution of this equation is given by
\begin{equation}
  \psi_n(q,t;\xi) = {\cal N}_n(q,t) P_n(q,t;\xi) E_{q^2}\left(\frac{t}{(q+1)q^n}
  \xi^2\right). \label{eq:es}
\end{equation}
Here $P_n(q,t;\xi)$ denotes an $n$th-degree polynomial in $\xi$, satisfying the relation
\begin{equation}
  P_{n+1}(q,t;\xi) = \xi P_n\left(q, \frac{t}{q}; \xi\right) - \xi \frac{t^2}{q^{n+1}}
  P_n\left(q, \frac{t}{q}; q\xi\right) - t {\cal D}_q P_n\left(q, \frac{t}{q}; \xi\right) 
  \label{eq:es-P}
\end{equation}
with $P_0(q,t;\xi) \equiv 1$, and ${\cal N}_n(q,t)$ is a normalization coefficient fulfilling
the recursion relation
\begin{equation}
  {\cal N}_{n+1}(q,t) = \left\{[n+1]_q \left(1
  - \frac{t^2}{q^{n+1}}\right)\right\}^{-1/2} {\cal N}_n\left(q, \frac{t}{q}\right) 
  \label{eq:es-N}
\end{equation}
with ${\cal N}_0(q,t)$ given in (\ref{eq:gs-norm}).\par
%
%
The solution of equation (\ref{eq:es-N}) is easily found to be given by
\begin{equation}
  {\cal N}_n(q,t) = \left\{[n]_q! (q^{-2n+1} t^2; q)_n\, {}_1\phi_0(q; \mbox{---}; q^2,
  q^{-2n} t^2)\right\}^{-1/2}
\end{equation}
where the symbol $(a; q)_n$ is defined in equation (\ref{eq:a-q}).\par
%
%
Solving equation (\ref{eq:es-P}) for the polynomials $P_n(q,t;\xi)$ looks however more
involved. First of all, let us remark that this equation may be considered as some
$q$-difference analogue of a differential equation satisfied by Hermite polynomials
\begin{equation}
  H_{n+1}(x) = 2x H_n(x) - H'_n(x)  \label{eq:hermite}
\end{equation}
which can be obtained by combining equations (22.7.13) and (22.8.7)
of~\cite{abramowitz}. As a matter of fact, in the formal limit where
$q\to1$ while $t^2$ remains different from one\footnote{By formal limit, we mean a limit
that does not correspond to any physical values of the parameters $\alpha$, $\beta$.
When $q\to1$, we indeed obtain from (\ref{eq:q}) that either $\alpha \to 0$ or $\beta
\to 0$, which, from (\ref{eq:k}), (\ref{eq:gamma}), and (\ref{eq:SI-cond2b}), implies that
either $t \to -1$ or $t \to 1$.}, equation (\ref{eq:es-P}) can be rewritten as
\begin{equation}
  Q_{n+1}(t;\xi) = (1-t^2) \xi Q_n(t;\xi) - t Q'_n(t;\xi)  \label{eq:hermite-bis}
\end{equation}
with $Q_n(t;\xi) \equiv P_n(1,t;\xi)$, because the $q$-differential operator ${\cal
D}_q$ goes to the ordinary differential operator $d/d\xi$. From (\ref{eq:hermite}), the
solution of (\ref{eq:hermite-bis}) is found to be
\begin{equation}
  Q_n(t;\xi) = c^n(t) H_n[a(t) \xi] \qquad a(t) = \sqrt{\frac{1-t^2}{2t}} \qquad c(t) =
  \sqrt{\frac{1}{2} t (1-t^2)}.
\end{equation}
\par
%
%
We may therefore regard $P_n(q,t;\xi)$ as some two-parameter deformation of Hermite
polynomials. As far as we know, such a deformation has not been considered elsewhere.
In the remainder of this section, we shall therefore devote ourselves to deriving an
explicit solution to equation (\ref{eq:es-P}).\par
%
%
It is straightforward to show that for the first few $n$ values, the polynomials
$P_n(q,t;\xi)$ are given by
\begin{eqnarray}
  P_1(q,t;\xi) & = & \left(1 - \frac{t^2}{q}\right) \xi  \label{eq:P-1} \\
  P_2(q,t;\xi) & = & \left(1 - \frac{t^2}{q^3}\right) \left[\left(1 - \frac{t^2}{q}\right)
        \xi^2 - t\right] \\
  P_3(q,t;\xi) & = & \left(1 - \frac{t^2}{q^5}\right) \left(1 - \frac{t^2}{q^3}\right)
        \left[\left(1 - \frac{t^2}{q}\right) \xi^3 - \frac{t}{q} (1+q+q^2) \xi\right]. 
        \label{eq:P-3}
\end{eqnarray}
\par
%
%
{}For general $n$ values, it is clear from the structure of equation (\ref{eq:es-P}) that
\begin{equation}
  P_n(q,t;-\xi) = (-1)^n P_n(q,t;\xi)
\end{equation}
which actually agrees with the examples shown in (\ref{eq:P-1}) -- (\ref{eq:P-3}). We
may therefore look for a solution  of the type
\begin{equation}
  P_n(q,t;\xi) = \sum_{m=0}^n  \case{1}{2}[1 + (-1)^{n-m}] f_{n,m}(q,t) \xi^m 
  \label{eq:ansatz}
\end{equation}
where $f_{n,m}(q,t)$, $m=n$, $n-2$,~\ldots, 0(1), are some yet undetermined
coefficients.\par
%
%
Inserting (\ref{eq:ansatz}) in (\ref{eq:es-P}) and using the property ${\cal D}_q \xi^m =
[m]_q \xi^{m-1}$, we get the recursion relations
\begin{eqnarray}
  f_{n+1,m}(q,t) & = & \left(1 - \frac{t^2}{q^{n-m+2}}\right) f_{n,m-1}\left(q, \frac{t}
        {q}\right) - t [m+1]_q\, f_{n,m+1}\left(q,\frac{t}{q}\right) \nonumber \\
  && m=n-1, n-3, \ldots, 0 (1)  \label{eq:recursion1} \\
  f_{n+1,n+1}(q,t) & = & \left(1 - \frac{t^2}{q}\right) f_{n,n}\left(q, \frac{t}{q}\right)
         \label{eq:recursion2}
\end{eqnarray}
with $f_{0,0}(q,t) \equiv 1$. In (\ref{eq:recursion1}), we have assumed that
$f_{n,-1}(q,t) = 0$ for odd values of $n$.\par
%
%
The solution of equation (\ref{eq:recursion2}) is given by
\begin{equation}
  f_{n,n}(q,t) = \prod_{k=0}^{n-1} \left(1 - \frac{t^2}{q^{2k+1}}\right) =
  \left(\frac{t^2}{q^{2n-1}}; q^2\right)_n  \label{eq:sol1}
\end{equation}
with $\prod_{k=0}^{-1} \equiv 1$ and $(a; q^2)_n$ defined as in (\ref{eq:a-q}). As
proved in the appendix, the solution of equation (\ref{eq:recursion1}) is provided by
\begin{equation}
  f_{n,m}(q,t) = \frac{[n]_q!}{[m]_q!\, [n-m]_q!!} \left(- \frac{t}{q^{(n+m-2)/2}}\right)
  ^{(n-m)/2} \left(\frac{t^2}{q^{2n-1}}; q^2\right)_{(n+m)/2}  \label{eq:sol2} 
\end{equation}
where $m = n-2$, $n-4$, ~\ldots, 0(1). Comparison between equations (\ref{eq:sol1})
and (\ref{eq:sol2}) shows that the latter can be extended to all allowed $m$ values, i.e.,
$m=n$, $n-2$,~\ldots, 0(1). Note that the property $f_{n,-1}(q,t) = 0$, for odd $n$
values, can be retrieved by using the usual assumption $[n]_q! \to \infty$ if $n \to
-1$.\par
%
%
Equations (\ref{eq:ansatz}) and (\ref{eq:sol2}) therefore provide us with the general
solution to equation (\ref{eq:es-P}), corresponding to $P_0(q,t;\xi) = 1$, and they include
equations (\ref{eq:P-1}) -- (\ref{eq:P-3}) as special cases.\par
%
%
To relate the polynomials $P_n(q,t;\xi)$ with some known basic polynomials, let us first
express them as basic hypergeometric functions~\cite{gasper}. Distinguishing between
even and odd $n$ values and using equations (\ref{eq:hyper}) -- (\ref{eq:rel-a-q}), we
obtain
\begin{eqnarray}
  P_{2\nu}(q,t;\xi) & = & (q; q^2)_{\nu} \left(\frac{t^2}{q^{4\nu-1}}; q^2\right)_{\nu}
         \left(\frac{t}{q^{\nu-1}(q-1)}\right)^{\nu} \nonumber \\
  && \mbox{} \times {}_2\phi_1\left[q^{-2\nu}, 
         \frac{t^2}{q^{2\nu-1}}; q; q^2, -q^{2\nu}(q-1) \frac{\xi^2}{t}\right]
         \label{eq:P-even} \\
  P_{2\nu+1}(q,t;\xi) & = & (q^3; q^2)_{\nu} \left(\frac{t^2}{q^{4\nu+1}};
         q^2\right)_{\nu+1} \left(\frac{t}{q^{\nu}(q-1)}\right)^{\nu} \xi \nonumber \\
  && \mbox{} \times {}_2\phi_1\left[q^{-2\nu}, 
         \frac{t^2}{q^{2\nu-1}}; q^3; q^2, -q^{2\nu+1}(q-1) \frac{\xi^2}{t}\right]
         \label{eq:P-odd}
\end{eqnarray}
where $\nu=0$, 1, 2,~\ldots. Since in (\ref{eq:P-even}) and (\ref{eq:P-odd}), the
parameter $a_1$ of the basic hypergeometric series is $a_1 = q^{-2\nu}$, where
$2\nu$ is a nonnegative integer, we deal here with terminating series. \par
%
%
{}From equations (\ref{eq:P-even}), (\ref{eq:P-odd}), and the definition of little
$q$-Jacobi polynomials~\cite{gasper}
\begin{equation}
  p_n(x; a, b; q) = {}_2\phi_1(q^{-n}, abq^{n+1}; aq; q, qx)
\end{equation}
it is now obvious that the polynomials $P_n(q,t;\xi)$ can be re-expressed in terms of the
latter as
\begin{eqnarray}
  P_{2\nu}(q,t;\xi) & = & (q; q^2)_{\nu} \left(\frac{t^2}{q^{4\nu-1}}; q^2\right)_{\nu}
         \left(\frac{t}{q^{\nu-1}(q-1)}\right)^{\nu} \nonumber \\
  && \mbox{} \times p_{\nu}\left[-q^{2\nu-2}(q-1) \frac{\xi^2}{t}; \frac{1}{q},
         \frac{t^2}{q^{4\nu}}; q^2\right] \\
  P_{2\nu+1}(q,t;\xi) & = & (q^3; q^2)_{\nu} \left(\frac{t^2}{q^{4\nu+1}};
         q^2\right)_{\nu+1} \left(\frac{t}{q^{\nu}(q-1)}\right)^{\nu} \xi \nonumber \\
  && \mbox{} \times p_{\nu}\left[-q^{2\nu-1}(q-1) \frac{\xi^2}{t}; q,
         \frac{t^2}{q^{4\nu+2}}; q^2\right].
\end{eqnarray}
\par
%
%
This completes the determination of the eigenfunctions (\ref{eq:es}) of the harmonic
oscillator Hamiltonian (\ref{eq:h}) in ($q$-deformed) Bargmann representation. Combining
equations (\ref{eq:phi-n}), (\ref{eq:gs}), (\ref{eq:es}), and (\ref{eq:ansatz}), we can
rewrite them as linear combinations of even- or odd-$n$-$q$-boson wave functions
\begin{eqnarray}
  \psi_{2\nu}(q,t;\xi) & = & {\cal N}_{2\nu}(q,t) \sum_{\sigma=0}^{\infty} \left[
         \sum_{\mu=0}^{\min(\sigma,\nu)} \frac{\sqrt{[2\sigma]_q!}}{[2\sigma -
         2\mu]_q!!} f_{2\nu,2\mu}(q,t)
         \left(\frac{t}{q^{2\nu}}\right)^{\sigma-\mu}\right] \nonumber \\
  && \mbox{} \times \varphi_{2\sigma}(q,\xi) \\
  \psi_{2\nu+1}(q,t;\xi) & = & {\cal N}_{2\nu+1}(q,t) \sum_{\sigma=0}^{\infty} \left[
         \sum_{\mu=0}^{\min(\sigma,\nu)} \frac{\sqrt{[2\sigma+1]_q!}}{[2\sigma -
         2\mu]_q!!} f_{2\nu+1,2\mu+1}(q,t)
         \left(\frac{t}{q^{2\nu+1}}\right)^{\sigma-\mu}\right] \nonumber \\
  && \mbox{} \times \varphi_{2\sigma+1}(q,\xi)
\end{eqnarray}
according to whether $n$ is even or odd.\par
%
%
\section{Conclusion}

In the present paper, we have determined in a purely algebraic way both the spectrum
and the eigenvectors of the harmonic oscillator with nonzero minimal uncertainties in
both position and momentum by availing ourselves of an extension of SUSYQM and shape
invariance powerful techniques to the case of the deformed canonical commutation
relation (\ref{eq:def-com}).\par
%
%
In the present context, shape invariance is related to the scaling of some parameter $t$,
depending in a complicated way upon the two deforming parameters $\aalpha$,
$\bbeta$ (or the two dimensionless ones $\alpha$, $\beta$) entering the canonical
commutation relation. As occurs in other examples involving parameter
scaling~\cite{spiridonov}, the oscillator spectrum turns out to be exponential. The
supersymmetric partner Hamiltonians correspond to both different masses and
frequencies. Such an unusual feature is a direct consequence of the deformation of the
commutation relation and is distinct from the combined potential and effective-mass
variations that may be effected in the case of a position-dependent effective mass in the
context of conventional SUSYQM~\cite{milanovic}.\par
%
%
We have proved that whenever one of the deforming parameters vanishes, e.g., $\alpha
\to 0$ and $\beta \ne 0$, our exponential spectrum goes to the quadratic one,
previously found by solving the deformed Schr\"odinger differential
equation~\cite{kempf95, chang}. Such a quadratic spectrum may also be derived from
SUSYQM and shape invariance connected with parameter translation.\par
%
%
{}Furthermore, we have shown that when $\alpha = \beta \ne 0$ or $\aalpha = m^2
\omega^2 \bbeta \ne 0$, the harmonic oscillator with nonzero minimal uncertainties in
both position and momentum reduces to the $q$-deformed harmonic oscillator
corresponding to $q>1$~\cite{cq} and its eigenvectors therefore coincide with the
$n$-$q$-boson states with $n=0$, 1, 2,~\ldots.\par
%
%
{}Finally, in the general case where $0 \ne \alpha \ne \beta \ne 0$, we have constructed
the oscillator eigenvectors as linear combinations of $n$-$q$-boson states by resorting to
a ($q$-deformed) Bargmann representation of the latter and to $q$-differential calculus.
The ground state can be expressed in terms of the $q$-exponential of an operator
proportional to the square of the $q$-boson creation operator, acting on the vacuum,
while the excited states contain as extra factors $n$th-degree polynomials in the
creation operator. The latter are some two-parameter deformations of Hermite polynomials
and can also be related to little $q$-Jacobi polynomials. It is worth noting that
operators similar, but not identical, to that occurring in the ground state eigenvector are
familiar in other contexts, such as those of squeezed states~\cite{dodonov} and of
Bose-Einstein condensates~\cite{navez}.\par
%
%
\section*{Acknowledgments}

One of the authors (CQ) would like to thank K A Penson for an interesting discussion. CQ
is a Research Director of the National Fund for Scientific Research (FNRS), Belgium.
VMT thanks this organization for financial support.\par
%
%
\section*{Appendix. Some results used in the determination of the harmonic oscillator
eigenvectors}
\renewcommand{\theequation}{A.\arabic{equation}}
\setcounter{section}{0}
\setcounter{equation}{0}

The $q$-exponential used in the present paper is defined by~\cite{exton}
\begin{equation}
  E_q(\xi) = \sum_{n=0}^{\infty} \frac{\xi^n}{[n]_q!}  
\end{equation}
where the $q$-factorial $[n]_q!$ is given in (\ref{eq:q-fac}). For $q>1$ (see equation
(\ref{eq:q})), it converges for all finite values of $\xi \in \C$. It is the solution of the
$q$-difference equation
\begin{equation}
  {\cal D}_q E_q(a\xi) = a E_q(a\xi) \qquad a \in \C  \label{eq:exp-eq}
\end{equation}
subject to the condition that $E_q(0) = 1$.\par
%
%
It follows from definition (\ref{eq:D-q}) and property (\ref{eq:exp-eq}) that
\begin{eqnarray}
  {\cal D}_q E_{q^2}(a \xi^2) & = & \frac{E_{q^2}(a q^2 \xi^2) - E_{q^2}(a \xi^2)}
          {(q-1)\xi} = a (q+1) \xi \frac{E_{q^2}(a q^2 \xi^2) - E_{q^2}(a \xi^2)}
          {(q^2-1) a \xi^2} \nonumber \\
  & = & a (q+1) \xi E_{q^2}(a \xi^2).  \label{eq:prop-exp}  
\end{eqnarray}
Hence the function $\psi_0(q,t;\xi)$, defined in (\ref{eq:gs}), satisfies the difference
equation (\ref{eq:gs-eq}).\par
%
%
The basic hypergeometric function ${}_r\phi_s(a_1, a_2, \ldots, a_r; b_1, b_2, \ldots,
b_s; q, z)$, generalizing the conventional one ${}_rF_s(a_1, a_2, \ldots, a_r; b_1, b_2,
\ldots, b_s; z)$, is defined by~\cite{gasper}
\begin{eqnarray}
  \lefteqn{{}_r\phi_s(a_1, a_2, \ldots, a_r; b_1, b_2, \ldots, b_s; q, z)\nonumber }\\
  & = & \sum_{n=0}^{\infty} \frac{(a_1;q)_n (a_2;q)_n \cdots (a_r;q)_n}{(q;q)_n
        (b_1;q)_n (b_2;q)_q \cdots (b_s;q)_n} [(-1)^n q^{n(n-1)/2}]^{1+s-r} z^n
        \label{eq:hyper}
\end{eqnarray} 
where $z \in \C$ and
\begin{equation}
  (a;q)_n  \equiv  \left\{\begin{array}{ll}
        1 & {\rm if\ } n=0 \\[0.2cm]
        (1-a) (1-aq) \cdots (1-aq^{n-1}) & {\rm if\ } n=1, 2, \ldots
     \end{array}\right. .  \label{eq:a-q}
\end{equation}
In (\ref{eq:hyper}), it is assumed that the parameters $b_1$, $b_2$,~\ldots, $b_s$ are
such that the denominator factors in the terms of the series are never zero. Since
$(q^{-m};q)_n=0$ if $n=m+1$, $m+2$,~\ldots, an ${}_r\phi_s$ series terminates if
one of its numerator parameters is of the form $q^{-m}$ with $m=0$, 1, 2,~\ldots.
When dealing with nonterminating series, it is assumed that the parameters and the
variable are such that the series converges absolutely.\par
%
%
Some useful relations connecting $(a;q)_n$ with the $q$-double factorials defined in
(\ref{eq:q-d-fac1}) and (\ref{eq:q-d-fac2}), as well as the latter with $q$-factorials, are
\begin{equation}
  [2\nu]_q!! = \frac{(q^2;q^2)_{\nu}}{(1-q)^{\nu}} \qquad [2\nu-1]_q!! =
  \frac{(q;q^2)_{\nu}}{(1-q)^{\nu}} \qquad [\nu]_{q^2}! = \frac{[2\nu]_q!!}
  {(q+1)^{\nu}}. 
\end{equation}
We may also note the interesting relations~\cite{gasper}
\begin{equation}
  (a;q)_{n+k} = (a;q)_n (a q^n;q)_k \qquad (a^{-1} q^{1-n};q)_n = (a;q)_n (-a^{-1})^n
  q^{-n(n-1)/2}.  \label{eq:rel-a-q}
\end{equation}
\par
%
%
Let us now prove that the functions $\psi_n(q,t;\xi)$, defined in (\ref{eq:es}), solve
equation (\ref{eq:es-eq}) provided conditions (\ref{eq:es-P}) and (\ref{eq:es-N}) are
satisfied. For such a purpose, we shall apply the well-known rule for $q$-derivating a
product of functions~\cite{exton}
\begin{equation}
  {\cal D}_q f(\xi) g(\xi) = f(q\xi) {\cal D}_qg(\xi) + g(\xi) {\cal D}_q f(\xi).
  \label{eq:product}
\end{equation}
Substituting equation (\ref{eq:es}) into the right-hand side of (\ref{eq:es-eq}) and using
(\ref{eq:product}) and (\ref{eq:prop-exp}) successively, we obtain
\begin{eqnarray}
  \lefteqn{\psi_{n+1}(q,t;\xi) \nonumber } \\ 
  & = & \left\{[n+1]_q \left(1 - \frac{t^2}{q^{n+1}}\right)\right\}^{-1/2} {\cal
        N}_n(q,t_1) \Biggl\{\xi P_n(q,t_1;\xi) E_{q^2}\left(\frac{t_1}{(q+1)q^n}
        \xi^2\right) \nonumber \\
  && \mbox{} - t P_n(q,t_1;q\xi) {\cal D}_q E_{q^2}\left(\frac{t_1}{(q+1)q^n}
        \xi^2\right) - t E_{q^2}\left(\frac{t_1}{(q+1)q^n} \xi^2\right) {\cal D}_q
        P_n(q,t_1;\xi) \Biggr\} \nonumber \\
  & = & \left\{[n+1]_q \left(1 - \frac{t^2}{q^{n+1}}\right)\right\}^{-1/2} {\cal
        N}_n(q,t_1) \Biggl\{\xi P_n(q,t_1;\xi) - t P_n(q,t_1;q\xi) \frac{t_1}{q^n} \xi
        \nonumber \\  
  && \mbox{}  - t {\cal D}_q
        P_n(q,t_1;\xi) \Biggr\} E_{q^2}\left(\frac{t_1}{(q+1)q^n} \xi^2\right) 
\end{eqnarray}
which should coincide with equation (\ref{eq:es}) with $n+1$ substituted for $n$.
Comparison between the right-hand sides of both equations directly leads to
(\ref{eq:es-P}) and (\ref{eq:es-N}), which completes the proof.\par
%
%
Let us finally consider the solution to the recursion relation (\ref{eq:recursion1}).
Substituting equation (\ref{eq:sol2}), where use is made of definition (\ref{eq:a-q}), into
the right-hand side of (\ref{eq:recursion1}), we get
\begin{eqnarray}
  \lefteqn{f_{n+1,m}(q,t) \nonumber} \\
  & = & \left(1 - \frac{t^2}{q^{n-m+2}}\right) \frac{[n]_q!}{[m-1]_q!\, [n-m+1]_q!!}
        \left(-\frac{t}{q^{(n+m-1)/2}}\right)^{(n-m+1)/2} \nonumber \\
  && \mbox{} \times \prod_{k=(n-m+1)/2}^{n-1} \left(1 - \frac{t^2}{q^{2k+3}}\right)
        \nonumber \\
  && \mbox{} - t [m+1]_q \frac{[n]_q!}{[m+1]_q!\, [n-m-1]_q!!}
        \left(-\frac{t}{q^{(n+m+1)/2}}\right)^{(n-m-1)/2} \prod_{k=(n-m-1)/2}^{n-1}
        \left(1 - \frac{t^2}{q^{2k+3}}\right) \nonumber \\  
  & = & \frac{[n]_q!}{[m]_q!\, [n-m+1]_q!!} ([m]_q + q^m [n-m+1]_q)
        \left(-\frac{t}{q^{(n+m-1)/2}}\right)^{(n-m+1)/2} \nonumber \\
  && \mbox{} \times \prod_{k=(n-m-1)/2}^{n-1} \left(1 - \frac{t^2}{q^{2k+3}}\right)
        \nonumber \\
  & = & \frac{[n+1]_q!}{[m]_q!\, [n-m+1]_q!!} 
        \left(-\frac{t}{q^{(n+m-1)/2}}\right)^{(n-m+1)/2} \left(\frac{t^2}{q^{2n+1}};q^2
        \right)_{(n+m+1)/2}  \label{eq:f}
\end{eqnarray}
where, in the last step, we employ the relation
\begin{equation}
  [m]_q + q^m [n-m+1]_q = [n+1]_q
\end{equation}
and equation (\ref{eq:a-q}) again. The final result in (\ref{eq:f}) coincides with equation
(\ref{eq:sol2}) for $n$ replaced by $n+1$. We conclude that equation (\ref{eq:sol2})
provides us with the solution to equation (\ref{eq:recursion1}) corresponding to
$f_{0,0}(q,t)=1$.\par
%
%


\newpage
\begin{thebibliography}{99}

\bibitem{gross} Gross D J  and Mende P F 1988 {\sl Nucl.\ Phys.} B {\bf 303} 407

\bibitem{maggiore} Maggiore M 1993 {\sl Phys.\ Lett.} B {\bf 304} 65

\bibitem{kempf94a} Kempf A 1994 Quantum field theory with nonzero minimal
uncertainties in position and momentum {\sl Preprint} hep-th/9405067

\bibitem{kempf94b} Kempf A 1994 {\sl J.\ Math.\ Phys.} {\bf 35} 4483

\bibitem{hinrichsen} Hinrichsen H and Kempf A 1996 {\sl J.\ Math.\ Phys.} {\bf 37} 2121

\bibitem{kempf97} Kempf A 1997 {\sl J.\ Phys.\ A: Math.\ Gen.} {\bf 30} 2093\\
Brau F 1999 {\sl J.\ Phys.\ A: Math.\ Gen.} {\bf 32} 7691

\bibitem{kempf93} Kempf A 1993 {\sl J.\ Math.\ Phys.} {\bf 34} 969

\bibitem{kempf95} Kempf A, Mangano G and Mann R B 1995 {\sl Phys.\ Rev.} D {\bf 52}
1108

\bibitem{chang} Chang L N, Minic D, Okamura N and Takeuchi T 2002 {\sl Phys.\ Rev.} D
{\bf 65} 125027

\bibitem{dadic} Dadi\'c I, Jonke L and Meljanac S 2003 {\sl Phys.\ Rev.} D {\bf 67}
087701

\bibitem{cooper} Cooper F, Khare A and Sukhatme U 1995 {\sl Phys.\ Rep.} {\bf 251}
267\\
Cooper F, Khare A and Sukhatme U 2001 {\sl Supersymmetry in Quantum Mechanics}
(Singapore: World Scientific)

\bibitem{junker} Junker G 1996 {\sl Supersymmetric Methods in Quantum and Statistical
Physics} (Berlin: Springer)

\bibitem{gendenshtein} Gendenshtein L E 1983 {\sl Pis'ma Zh.\ Eksp.\ Teor.\ Fiz.} {\bf
38} 299 [{\sl JETP Lett.} {\bf 38} 356]

\bibitem{dabrowska} Dabrowska J, Khare A and Sukhatme U 1988 {\sl J.\ Phys.\ A:
Math.\ Gen.} {\bf 21} L195

\bibitem{schrodinger} Schr\"odinger E 1940 {\sl Proc.\ R.\ Irish Acad.} A {\bf 46} 9,
183 \\
Schr\"odinger E 1941 {\sl Proc.\ R.\ Irish Acad.} A {\bf 47} 53

\bibitem{infeld} Infeld L and Hull T E 1951 {\sl Rev.\ Mod.\ Phys.} {\bf 23} 21

\bibitem{spiridonov} Spiridonov V 1992 {\sl Phys.\ Rev.\ Lett.} {\bf 69} 398 \\
Spiridonov V 1992 {\sl Mod.\ Phys.\ Lett.} A {\bf 7} 1241

\bibitem{khare} Khare A and Sukhatme U P 1993 {\sl J.\ Phys.\ A: Math.\ Gen.} {\bf 26}
L901 \\
Barclay D T, Dutt R, Gangopadhyaya A, Khare A, Pagnamenta A and Sukhatme U 1993
{\sl Phys.\ Rev.} A {\bf 48} 2786

\bibitem{green} Green H S 1965 {\sl Matrix Mechanics} (Groningen: P.\ Noordhoff LTD)

\bibitem{milanovic} Milanovi\'c V and Ikoni\'c Z 1999 {\sl J.\ Phys.\ A: Math.\ Gen.} {\bf
32} 7001

\bibitem{arik} Arik M and Coon D D 1976 {\sl J.\ Math.\ Phys.} {\bf 17} 524

\bibitem{cq} Quesne C, Penson K A and Tkachuk V M 2003 {\sl Phys.\ Lett.} A {\bf 313}
29

\bibitem{exton} Exton H 1983 {\sl $q$-Hypergeometric Functions and Applications}
(Chichester: Ellis Horwood)

\bibitem{gasper} Gasper G and Rahman M 1990 {\sl Basic Hypergeometric Series}
(Cambridge: Cambridge University Press)

\bibitem{abramowitz} Abramowitz M and Stegun I A 1965 {\sl Handbook of Mathematical
Functions} (New York: Dover)

\bibitem{dodonov} Dodonov V V 2002 {\sl J.\ Opt.\ B: Quantum Semiclass.\ Opt.} {\bf
4} R1

\bibitem{navez} Navez P 1998 {\sl Mod.\ Phys.\ Lett.} B {\bf 12} 705\\
Solomon A I, Feng Y and Penna V 1999 {\sl Phys.\ Rev.} B {\bf 60} 3044

\end {thebibliography}
\end{document}